\documentclass[proof]{WileyASNA-v1}
\usepackage{supertabular}
\usepackage{lscape}
\articletype{Article Type}%

\received{03 October 2020}
\revised{28 October 2020}
\accepted{09 November 2020}

\begin{document}

\title{Spectroscopic orbit determination of the long-periodic binary system $\theta$\,Cep\protect\thanks{Based on observations obtained with telescopes of the University Observatory Jena, which is operated by the Astrophysical Institute of the Friedrich-Schiller University.}}

\author[]{Richard Bischoff*}
\author[]{Markus Mugrauer}
\author[]{Oliver Lux}
\author[]{Tamara Zehe}
\author[]{Therese Heyne}
\author[]{Daniel Wagner}
\author[]{Michael Geymeier}

\authormark{Bischoff \textsc{et al.}}

\address[]{\orgdiv{Astrophysikalisches Institut und Universit\"{a}ts-Sternwarte Jena},  \orgaddress{\state{Schillerg\"{a}{\ss}chen 2, 07745 Jena}, \country{Germany}}}

\corres{*R. Bischoff, Astrophysikalisches Institut und Universit\"{a}ts-Sternwarte Jena, Schillerg\"{a}{\ss}chen 2, 07745 Jena, Germany \email{richard.bischoff@uni-jena.de}}

\abstract{In 2015 a radial velocity monitoring campaign was started in order to redetermine and/or constrain the orbital solutions of spectroscopic binary systems. The observations were carried out at the University Observatory Jena with the \'{E}chelle spectrograph FLECHAS. The results from the main part of our target sample are already published. For the final target of this campaign, $\theta$\,Cep, we can now present an orbital solution based on a homogeneously covered radial velocity curve. The period of this single-lined spectroscopic binary turns out to be significantly larger and the orbit is much more eccentric compared to the given values in the \textit{9th Catalogue of Spectroscopic Binary Orbits}.}



\maketitle

\section{Introduction}\label{sec1}

From February 2015 until September 2020, a radial velocity ($RV$) monitoring campaign of selected binary systems was carried out at the University Observatory Jena, which is located close to the small village Gro{\ss}schwabhausen \citep{pfau}. The 90-cm reflector was operated in Nasmyth mode ($D=90$\,cm, $f/D=15$) for the spectroscopic observations. The telescope is equipped with the fiber-linked \'{E}chelle spectrograph FLECHAS \citep{mugrauer2014}. The aim of this campaign was to record $RV$ data of spectroscopic binaries and to constrain or improve their orbital solutions. The main part of our target sample are already published in \cite{bischoff} and \cite{heyne}. The $RV$ measurements and orbital solution of $\theta$\,Cep, presented in this paper, are the completion of our monitoring campaign.

All targets from our original sample were chosen from the \textit{9th Catalogue of Spectroscopic Binary Orbits} (SB9 hereafter) from \cite{pourbaix}. The SB9 catalogue is available in the \texttt{VizieR}\footnote{\url{https://cdsarc.unistra.fr/viz-bin/cat/B/sb9}} database \citep{ochsenbein} and contains information about orbital elements of several thousand spectroscopic binary stars. The precision of the orbital solutions in the SB9 is represented by different orbit grades. The grades range from 0 (poor) to 5 (definitive).
$\theta$\,Cep had a grade of 1 and was therefore suitable as a target for our observations. Furthermore, it is bright enough to record FLECHAS spectra with sufficiently high signal-to-noise-ratio (SNR) $>100$ within a few minutes of integration time and it is observable at air masses $X<2.5$ throughout the year. The original orbital elements of $\theta$\,Cep are listed in Table\,\ref{ueberblick} together with further information from the SB9 catalogue.

\begin{table*}[h!]
\caption{Properties of $\theta$\,Cep: right ascension $RA$, declination $Dec$, visual magnitude $V$, spectral type SpT, and the orbital elements of the system, namely its orbital period $P$, periastron time $T_{0}$, $RV$ semi-amplitude $K$, systemic velocity $\gamma$, orbital eccentricity $e$ and argument of periastron $\omega$ as given in the SB9.}
\centering
\begin{tabular}{cccccccccc}
\hline
$RA$       & $Dec$      &  $V$    &   SpT & $P$  & $T_{0}$  & $K$    & $\gamma$  & $e$ & $\omega$\\
	
[hh mm ss.ss] &  [dd mm ss.s] &   [mag] &       &  [d] & [JD]     & [km/s] & [km/s]    &     &  [$^{\circ}$]    \\
\hline
20 29 34.89 & +62 59 38.8	& 4.21 & Am & 840.6 & 2416214.5 & 13.85 & -6.4 & 0.03 & 83.7  \\
\hline
\end{tabular}
\begin{flushleft}
The orbital elements were taken from \cite{abt1961}.
\end{flushleft}    		
\label{ueberblick}
\end{table*}

In section 2 we explain the observations and data reduction of our spectra. The following section contains the description of the $RV$ measurements. Section 4 yields the determination of the orbital solution of $\theta$\,Cep. The results are discussed in the final section.

\section{Observations and data reduction}\label{sec2}

During four observing epochs between February 2016 and September 2020, $\theta$\,Cep was observed 197 times with FLECHAS at the University Observatory Jena during our $RV$ monitoring campaign. A detailed chronological overview about the different observing epochs is given in Table\,\ref{obslog}\hspace{-2mm}. These observations contain a total integration time of 24.6\,h.

\begin{table}[h!]
	\caption{Observation log. We list for each observing epoch the number of observations $N_{\text{Obs}}$, and the associated time span.}
	\centering
	\begin{tabular}{ccll}
		\hline
		Epoch      & $N_{\text{Obs}}$   &  begin & end       \\
		\hline
		1 & 41 & Feb 12, 2016 & Sep 27, 2016 \\
		2 & 24 & Dec 05, 2016 & May 06, 2017 \\
		3 & 50 & Jan 27, 2018 & Jun 03, 2018 \\
		4 & 82 & Mar 03, 2020 & Sep 21, 2020 \\
		\hline
	\end{tabular}                                              		
	\label{obslog}
\end{table}

FLECHAS covers the spectral range from  3900\,\AA\,\,to 8100\,\AA\,\,in 29 orders with a resolving power of $R\approx9300$. The instrument is equipped with a back-illuminated CCD-sensor (E2V CCD47-10), which consists of 1056\,x\,1027\,pixels, each a square with an edge length of 13\,$\mu$m. An overscan region is always read out to measure and correct the bias level. The typical read-noise of the FLECHAS detector is about 11\,$e^{-}$ and the gain is $1.3\,e^{-}$/ADU. Further details on the FLECHAS CCD-sensor are described in \cite{mugrauer2014}.

The observations of $\theta$\,Cep were carried out in the 1x1 binning mode with an individual integration time of 150\,s per spectrum. At each observation, three target spectra were recorded, to remove cosmics in the individual spectra and to guarantee a high SNR. This results in an average SNR of 253 as measured in at 6520\,\AA\ for all observations. The individual SNR range between 119 and 438 in all fully reduced spectra.

Before each observation of $\theta$\,Cep, three frames of a tungsten lamp and three frames of ThAr lamp were recorded for flat-fielding and wavelength calibration, respectively. The individual integration time of each calibration image is 5\,s. \cite{irrgang}, \cite{bischoff} and \cite{heyne} have already shown the long-term stability of the wavelength calibration of FLECHAS. For dark current and bias subtraction, in each observing night, three dark frames were recorded for the used integration times.

The calibration of our data was performed with the FLECHAS pipeline. This software was developed at the Astrophysical Institute Jena and includes dark current subtraction, flat-fielding, the extraction and wavelength calibration of the individual spectral orders, as well as averaging and normalisation of the obtained spectra \citep{mugrauer2014}.

\section{Radial velocity measurements}\label{sec3}

For $RV$ measurements of $\theta$\,Cep, we determined the central wavelengths $\lambda$ of the Balmer lines $H_{\alpha}$ ($\lambda_0=6562.81$\,\AA), $H_{\beta}$ ($\lambda_0=4861.34$\,\AA), and $H_{\gamma}$ ($\lambda_0=4340.48$\,\AA). This procedure is consistent with the previous work of our $RV$ monitoring campaign presented by \cite{bischoff} and \cite{heyne}. Furthermore, the hydrogen lines are the most prominent of the A7\,III star \citep{Kharchenko}, which is the main component of this binary.

The determination of line centres was performed with the IRAF script \texttt{splot}, which includes Gaussian fitting for the Doppler broadened cores. The measured wavelength shifts $\lambda - \lambda_0$ contain the $RV$ of $\theta$\,Cep, namely

\begin{equation}
RV = c \cdot \dfrac{\lambda-\lambda_0}{\lambda_0}+BC,
\end{equation}

where $c$ is the speed of light and $BC$ the barycentric correction. This correction takes the $RV$ offset between the observing site and the barycentre of the solar system into account. Furthermore, each date of observation was converted from Julian Date into Barycentric Julian Date (BJD). The derived average $RV$s of all observing epochs are listed in Table\,\ref{tab:RV}\hspace{-2mm}. Our achieved precision for the $RV$ measurements is 0.38\,km/s.

In addition, we also checked the stability of the wavelength calibration of the instrument by measuring telluric lines in all spectra. The average wavelength scatter as measured from the telluric lines is ($0.00\pm0.55$)\,km/s and is therefore consistent with the achieved $RV$ precision. No strong deviations or trends were found within the calibration during the whole $\theta$\,Cep campaign spanning $\sim4.6$\,yr of spectroscopic observations.

\section{Orbit determination}\label{sec4}

The Keplerian orbital elements, namely the semi-amplitude $K$, the eccentricity $e$, the argument of periastron $\omega$, the true anomaly $\nu$ and the systemic velocity $\gamma$, characterise the changing $RV$ of the primary component in a single-lined spectroscopic binary system. This periodic change can be described by
\begin{equation}
RV = K\cdot\left[e\cdot\cos(\omega)+\cos(\nu+\omega) \right]+\gamma,
\end{equation}
and represents the motion of the primary around the barycentre of the system. In addition, the semi-amplitude $K$ also contains the information about the orbital Period $P$ and the minimum semi-major axis $a\cdot\sin(i)$ via
\begin{equation}
K= \frac{2\pi\cdot a\sin(i)}{P\cdot\sqrt{1-e^{2}}}.
\end{equation}
The Keplerian elements of $\theta$\,Cep were determined by fitting orbital solutions on our obtained $RV$ data by using the spectroscopic binary solver \citep{johnson}. The first step was to identify the orbital period. Therefore, we chose the normalised periodogram \citep{lomb}, the string length \citep{dworetsky}, and the phase-dispersion-minimisation method \citep{marraco} to derive the best fitting orbital period in three independent ways. All solutions of the different algorithms are consistent with each other and therefore, we adopted the average as initial period for the calculation of the remaining orbital elements. The final orbital solution and its uncertainties were then determined with the method of least squares. The results are presented in Table\,\ref{neu_ele}\hspace{-2mm}. The fitted and phase folded $RV$ curve of $\theta$\,Cep, which covers about 1.8 orbital periods, is illustrated in Fig.\ref{RV2}\hspace{-2mm}. We also show the individual measurements with their typical precision of 0.38\,km/s.

\begin{table*}[h!]
	\caption{The derived Keplerian orbital elements of $\theta$\,Cep with their uncertainties.}
	\centering
	\begin{tabular}{cccccc}
		\hline
		 $P$  & $T_{0}$                     & $K$    & $\gamma$  & $e$ & $\omega$\\
		
		  [d] & [BJD - 2450000]             & [km/s] & [km/s]    &     &  [$^{\circ}$]    \\
		\hline
		 $914.3\pm7.0$ & $6991.3\pm8.5$ & $7.72\pm0.06$ & $-4.14\pm0.04$ & $0.377\pm0.006$ & $48.8\pm1.1$  \\
		\hline
	\end{tabular} 		
	\label{neu_ele}
\end{table*}

\section{Discussion}\label{sec5}

During our monitoring campaign of the grade-1 system $\theta$\,Cep, in total 172 $RV$ measurements were carried out, which spans 1.8 orbital periods. We were able to redetermine its Keplerian elements, presented in Table\,\ref{neu_ele}\hspace{-2mm}, and found significant differences to those given in the SB9 catalogue. The orbital solution was determined with least-square fitting of spectroscopic orbits on the given $RV$ measurements. This exhibits a reduced chi-square value of $\chi^{2}_{\text{red}}=1.03$. The resulting phase-folded $RV$ curve is homogeneously covered with data points, that have uncertainties which are 20 times smaller compared to the derived semi-amplitude.

\begin{figure*}[h!]
	 \centering\includegraphics[width=17.5cm,height=17.5cm,keepaspectratio]{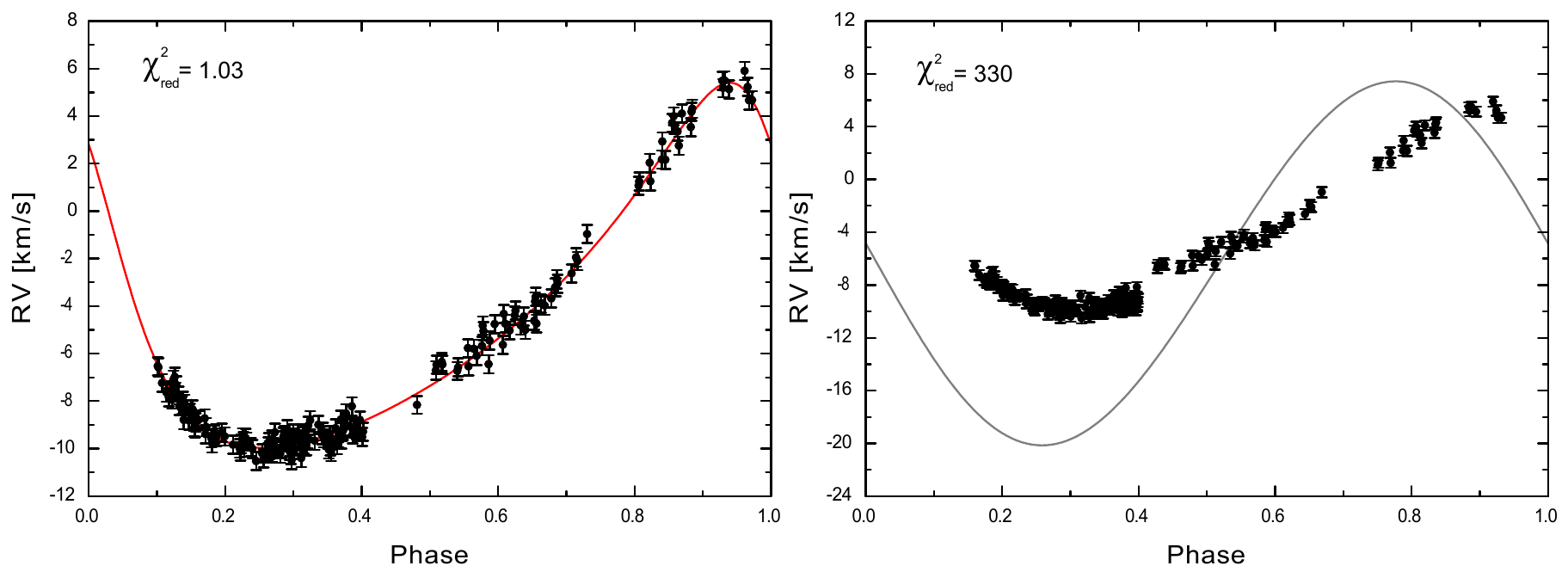}
	\caption{$RV$ measurements of $\theta$\,Cep, taken in the course of our spectroscopic monitoring campaign with FLECHAS. The individual $RV$ measurements are shown with their uncertainties together with the derived best fitting Keplerian orbital solution, shown as red line on the left. The orbital solution from the SB9 on the right (grey line), is clearly inconsistent with the $RV$ data, taken in this study.}
	\label{RV2}
\end{figure*}

As seen in Fig.\,\ref{RV2}\hspace{-2mm}, our measurements disagree with the orbital solution published by \cite{abt1961} and listed in the SB9 catalogue. The obtained orbital period of $\theta$\,Cep from our study is about $73.74$\,d longer. In addition, we determined a much more eccentric orbit in comparison to the nearly circular one given by \cite{abt1961}.

\begin{figure}[h!]
	 \centering\includegraphics[width=8.75cm,height=8.75cm,keepaspectratio]{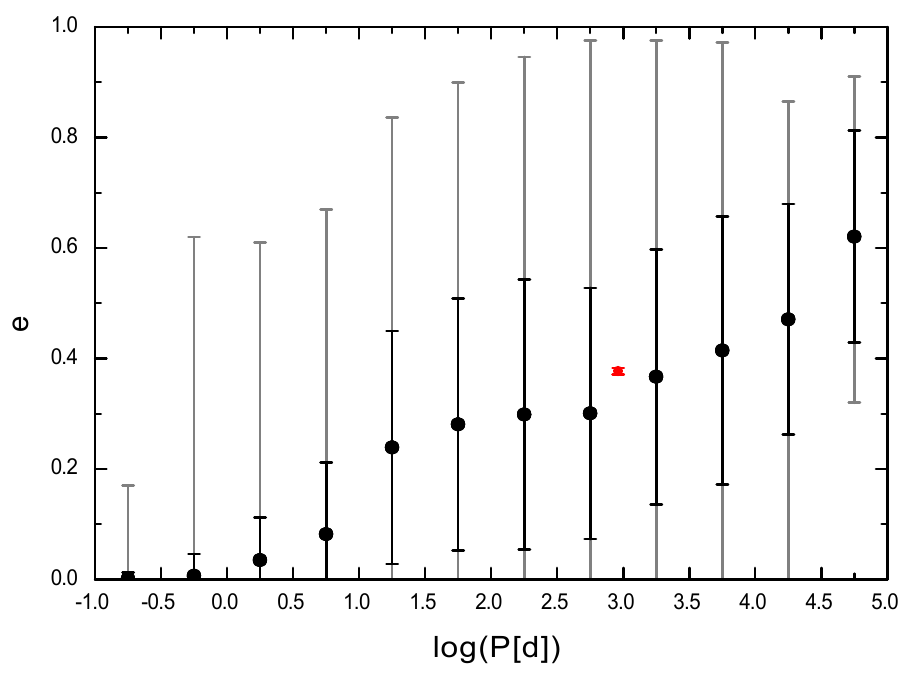}
	\caption{Average eccentricity for 5013 binary stars from the SB9 catalogue plotted against their binned logarithm of orbital periods (black dots). The standard deviations are shown as black bars and the complete ranges of each interval are illustrated as gray bars. $\theta$\,Cep is given as red dot with its uncertainties.}
	\label{meanecc}
\end{figure}

In contrast to this, our derived orbital solution fits very well with the period-eccentricity-correlation of spectroscopic binaries, as illustrated in Fig.\,\ref{meanecc}\hspace{-2mm}. $\theta$\,Cep is located near the average and within the standard deviation of the corresponding interval of periods.

Our found semi-amplitude is by a factor of 0.56 smaller compared to the corresponding value in the SB9. Furthermore, the systemic velocity differs by 2.26\,km/s and the argument of periastron by 34.9$^{\circ}$ from the orbital solution in the SB9.

 In the end, the mass-function
 \begin{equation}
 f(M)=\frac{M^{3}_{2}\cdot\sin^{3}(i)}{(M_{1}+M_{2})^{2}}=\frac{(1-e^{2})^{3/2}\cdot P\cdot K^{3}}{2\pi\cdot G}
 \end{equation}
 \newline
 and the minimum semi-major axis
 \begin{equation}
 a\cdot\sin(i)=\frac{P\cdot K\cdot \sqrt{1-e^{2}}}{2\pi}
 \end{equation}
\newline
 can be derived from the determined orbital elements, which are $f(M)=(35\pm1)\cdot10^{-3}$\,M$_{\odot}$ and $a\cdot\sin(i)=(601\pm 7)\cdot10^{-3}$\,au in the case of $\theta$\,Cep.

 Given its parallax of $25.7611\pm0.5808$\,mas from \textit{Gaia}\,DR2 \citep{gaia}, the expected astrometric orbit corresponds to $a\cdot\sin(i)=(15.6_{-0.5}^{+0.6})$\,mas. Therefore, it is possible that an orbital solution for $\theta$\,Cep can be determined with \textit{Gaia} data, which will be provided by a future data release.

 Furthermore, we can use the mass estimates for $\theta$\,Cep $M_{1}=(1.98_{-0.22}^{+0.25})$\,M$_{\odot}$ from the \textit{StarHorse} catalogue \citep{anders} to calculate the minimum mass of the secondary $M_{2}\cdot\sin(i)=(0.62\pm0.05)$\,M$_{\odot}$.

 By adopting that the companion is a dwarf, its derived minimum mass fits best with a K7\,V star with $R=(0.654_{-0.102}^{+0.044})$\,R$_{\odot}$ and $T_{\text{eff}}=(4070_{-130}^{+340})$\,K \citep{mamajek}\footnote{online available at \url{https://www.pas.rochester.edu/~emamajek/EEM_dwarf_UBVIJHK_colors_Teff.txt}}. Based on $T_{\text{eff}}$ from \textit{StarHorse} and the radius from \textit{Gaia}\,DR2 for the primary, this results in a luminosity ratio of $L_{1}/L_{2}=395_{-244}^{+747}$. Therefore, it is not possible to detect lines of the secondary within the spectra.

Our $RV$ monitoring campaign, which started in 2015, is now completed and shows how modern spectroscopic observations improve orbital solutions significantly and are necessary for correct statistics of binary systems.

The individual $RV$ measurements and derived orbital solution of $\theta$\,Cep, presented in this paper, will be available online in \texttt{VizieR}.

\section*{Acknowledgments}

We thank additional observers who have been involved in some of the observations of this project, obtained at the University Observatory Jena, in particular H. Gilbert, D. W\"{o}ckel, S. Hoffmann, V. Munz, W. Stenglein and K.-U. Michel.\\
This publication makes use of data products of the \texttt{VizieR} databases and \texttt{SIMBAD}, operated at CDS, Strasbourg, France.\\
This work was supported by the Deutsche Forschungsgemeinschaft with the projects \fundingNumber{NE 515/58-1} and \fundingNumber{MU 2695/27-1}.

\bibliography{bischoff}

\section*{Author Biography}

Richard Bischoff is a PhD student at the Astrophysical Institute and 
University Observatory Jena. His main fields of research are photometry
and spectroscopy of exoplanet candidate host stars.\vfill

\includegraphics[height=10cm]{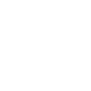}\vfill

\includegraphics[height=10cm]{empty.pdf}\vfill

\includegraphics[height=10cm]{empty.pdf}\vfill

\appendix

\section{Radial velocity measurements\label{app1}}

\begin{center}
\vspace{0.8cm}
\tablefirsthead{%
\hline
Date of Obs.          & \multicolumn{1}{c}{$RV$}  \\
BJD - 2450000         & \multicolumn{1}{c}{[km/s]}  \\
\hline}
\tablehead{%
\multicolumn{1}{l}{Continued}\\
\hline
Date of Obs.          & \multicolumn{1}{c}{$RV$} \\
BJD - 2450000         & \multicolumn{1}{c}{[km/s]} 	\\
\hline}
\tabletail{%
\hline	}
\tablelasttail{\hline}
\tablecaption{$RV$ measurements of $\theta$\,Cep. The typical $RV$ precision is 0.38\,km/s.}
\begin{supertabular}{cr}
7431.41999  & $-8.16  $ \\
7456.52214  & $-6.71  $ \\
7457.51626  & $-6.47  $ \\
7464.45784  & $-6.35  $ \\
7465.48188  & $-6.46  $ \\
7485.38406  & $-6.73  $ \\
7486.40238  & $-6.58  $ \\
7499.58034  & $-5.77  $ \\
7500.33442  & $-6.54  $ \\
7507.37131  & $-5.80  $ \\
7511.36688  & $-6.10  $ \\
7518.41141  & $-5.68  $ \\
7519.51279  & $-4.82  $ \\
7520.35886  & $-5.05  $ \\
7527.40683  & $-6.45  $ \\
7528.54019  & $-5.45  $ \\
7535.39983  & $-4.76  $ \\
7546.49179  & $-5.63  $ \\
7547.47012  & $-4.33  $ \\
7549.43703  & $-4.74  $ \\
7555.43857  & $-5.03  $ \\
7562.42892  & $-4.40  $ \\
7563.42764  & $-4.19  $ \\
7570.50335  & $-4.82  $ \\
7574.51102  & $-4.44  $ \\
7576.51069  & $-5.00  $ \\
7588.39623  & $-4.62  $ \\
7589.50490  & $-3.81  $ \\
7590.36447  & $-3.61  $ \\
7591.41527  & $-4.74  $ \\
7597.40810  & $-3.89  $ \\
7602.44804  & $-3.98  $ \\
7611.43795  & $-3.68  $ \\
7616.40631  & $-3.22  $ \\
7617.33948  & $-3.20  $ \\
7618.58584  & $-2.87  $ \\
7619.56147  & $-3.06  $ \\
7638.49600  & $-2.63  $ \\
7644.49182  & $-1.95  $ \\
7646.45344  & $-2.10  $ \\
7659.25949  & $-0.97  $ \\
7728.25429  & $1.07	  $ \\
7729.18617  & $1.25	  $ \\
7743.22034  & $2.03	  $ \\
7744.17682  & $1.25	  $ \\
7759.56783  & $2.17	  $ \\
7760.20534  & $2.93	  $ \\
7764.23047  & $2.16	  $ \\
7773.20956  & $3.71	  $ \\
7775.29681  & $3.99	  $ \\
7776.21782  & $3.63	  $ \\
7780.58626  & $3.35	  $ \\
7782.22415  & $2.75	  $ \\
7786.37238  & $4.10	  $ \\
7798.23398  & $3.53	  $ \\
7799.24356  & $4.18	  $ \\
7800.26635  & $4.31	  $ \\
7840.44945  & $5.48	  $ \\
7841.39590  & $5.18	  $ \\
7843.38752  & $5.50	  $ \\
7849.48172  & $5.13	  $ \\
7870.37340  & $5.90	  $ \\
7874.34512  & $5.22	  $ \\
7876.45511  & $4.66	  $ \\
7880.40963  & $4.67	  $ \\
8146.23746  & $-9.91  $ \\
8151.25549  & $-9.61  $ \\
8155.28913  & $-9.32  $ \\
8162.29514  & $-10.22 $ \\
8164.23534  & $-10.02 $ \\
8166.58936  & $-9.51  $ \\
8170.70087  & $-9.58  $ \\
8171.59772  & $-9.19  $ \\
8173.04483  & $-9.64  $ \\
8173.63903  & $-9.75  $ \\
8174.65618  & $-10.10 $ \\
8176.62476  & $-10.27 $ \\
8177.57345  & $-10.49 $ \\
8178.52010  & $-9.79  $ \\
8179.53878  & $-9.37  $ \\
8190.49365  & $-10.40 $ \\
8197.51657  & $-9.89  $ \\
8202.52572  & $-8.81  $ \\
8208.44996  & $-9.67  $ \\
8213.39842  & $-9.00  $ \\
8218.46306  & $-9.43  $ \\
8219.56977  & $-9.28  $ \\
8220.51333  & $-9.56  $ \\
8223.43816  & $-9.56  $ \\
8226.40497  & $-9.63  $ \\
8227.41988  & $-10.04 $ \\
8228.46644  & $-9.44  $ \\
8229.46712  & $-9.91  $ \\
8230.39943  & $-10.15 $ \\
8232.40048  & $-9.66  $ \\
8236.58244  & $-9.14  $ \\
8238.45940  & $-9.85  $ \\
8240.46288  & $-9.56  $ \\
8243.44872  & $-8.78  $ \\
8244.39035  & $-9.62  $ \\
8245.39480  & $-9.30  $ \\
8246.40793  & $-9.25  $ \\
8247.37977  & $-9.15  $ \\
8248.43526  & $-9.12  $ \\
8250.43803  & $-8.52  $ \\
8256.36127  & $-9.08  $ \\
8258.38285  & $-8.22  $ \\
8259.39365  & $-9.41  $ \\
8260.39249  & $-9.15  $ \\
8264.44846  & $-9.22  $ \\
8265.45145  & $-9.08  $ \\
8267.43412  & $-9.30  $ \\
8269.40108  & $-8.82  $ \\
8272.33986  & $-9.51  $ \\
8273.32986  & $-9.29  $ \\
8912.52884  & $-6.53  $ \\
8913.46328  & $-6.58  $ \\
8918.31542  & $-7.24  $ \\
8924.30676  & $-7.65  $ \\
8927.40105  & $-7.84  $ \\
8930.45148  & $-7.56  $ \\
8931.49983  & $-7.94  $ \\
8932.47465  & $-7.56  $ \\
8933.50080  & $-7.10  $ \\
8934.51247  & $-7.62  $ \\
8935.45552  & $-6.98  $ \\
8936.44046  & $-7.61  $ \\
8937.47338  & $-7.34  $ \\
8940.46748  & $-7.76  $ \\
8941.43652  & $-8.21  $ \\
8944.43397  & $-8.22  $ \\
8945.42498  & $-8.15  $ \\
8946.48005  & $-7.98  $ \\
8947.42760  & $-8.80  $ \\
8948.45568  & $-8.12  $ \\
8949.42870  & $-8.35  $ \\
8954.36481  & $-8.82  $ \\
8955.44782  & $-8.50  $ \\
8957.43669  & $-8.33  $ \\
8959.42789  & $-8.35  $ \\
8960.42241  & $-8.86  $ \\
8961.41281  & $-9.11  $ \\
8962.45251  & $-9.14  $ \\
8963.46044  & $-8.89  $ \\
8966.42891  & $-8.81  $ \\
8967.41995  & $-8.79  $ \\
8975.44447  & $-8.73  $ \\
8976.41666  & $-9.39  $ \\
8977.34096  & $-9.09  $ \\
8984.44897  & $-9.57  $ \\
8985.37152  & $-9.78  $ \\
8987.40375  & $-9.38  $ \\
8988.36148  & $-9.39  $ \\
8990.40409  & $-9.35  $ \\
8991.35863  & $-9.52  $ \\
8998.41012  & $-9.33  $ \\
9002.36343  & $-9.49  $ \\
9013.38198  & $-9.84  $ \\
9022.37428  & $-9.99  $ \\
9023.42449  & $-10.13 $ \\
9024.37176  & $-9.81  $ \\
9026.37179  & $-9.53  $ \\
9027.38129  & $-9.68  $ \\
9028.36772  & $-9.97  $ \\
9030.36482  & $-9.55  $ \\
9031.37769  & $-9.71  $ \\
9037.37313  & $-9.93  $ \\
9039.38033  & $-10.01 $ \\
9044.36936  & $-10.53 $ \\
9052.39858  & $-10.20 $ \\
9054.39103  & $-10.42 $ \\
9060.41157  & $-10.16 $ \\
9061.35545  & $-9.73  $ \\
9062.34965  & $-10.07 $ \\
9066.34688  & $-10.20 $ \\
9067.34899  & $-9.92  $ \\
9068.35533  & $-10.04 $ \\
9069.34684  & $-9.98  $ \\
9080.44678  & $-9.66  $ \\
9081.31853  & $-9.46  $ \\
9082.31924  & $-9.43  $ \\
9089.32311  & $-9.94  $ \\
9094.40772  & $-10.06 $ \\
9095.40337  & $-9.53  $ \\
9100.28569  & $-9.34  $ \\
9101.33557  & $-9.60  $ \\
9103.44026  & $-9.90  $ \\
9104.29013  & $-9.84  $ \\
9105.30253  & $-9.70  $ \\
9107.29556  & $-9.49  $ \\
9108.27288  & $-9.84  $ \\
9109.38978  & $-9.40  $ \\
9110.27192  & $-9.37  $ \\
9111.29569  & $-9.90  $ \\
9112.29624  & $-9.67  $ \\
9113.26578  & $-9.62  $ \\
9114.26467  & $-9.03  $ \\
\end{supertabular}
\label{tab:RV}
\end{center}

\end{document}